%% file: sc/main.tex
\documentclass[conference]{IEEEtran}

\usepackage{_cfg-paper}

\usepackage{flushend}
\usepackage{subfiles}
\usepackage{enumitem}
\usepackage{color, colortbl}
\usepackage{tikz}
\usepackage{caption} 
\usepackage{comment}
\usepackage{tabularx, color, colortbl}
\usepackage{hyperref}       % hyperlinks
\usepackage{url}            % simple URL typesetting
\usepackage{booktabs}       % professional-quality tables
\usepackage{amsfonts}       % blackboard math symbols
\usepackage{nicefrac}       % compact symbols for 1/2, etc.
\usepackage{microtype}      % microtypography
\usepackage{lipsum}
\usepackage{amsthm}
\usepackage{multirow}
\usepackage{bm}
\usepackage[normalem]{ulem}
\usepackage[vlined, ruled, linesnumbered]{algorithm2e}
\SetAlgoInsideSkip{0pt}      % try 0pt–3pt
\usepackage{subcaption}
\renewrobustcmd{\bfseries}{\fontseries{b}\selectfont}
\renewrobustcmd{\boldmath}{}
\newrobustcmd{\B}{\bfseries}
\definecolor{Gray}{gray}{0.9}
\newcommand*\circled[1]{\tikz[baseline=(char.base)]{
            \node[shape=circle,fill,inner sep=1pt] (char) {\textcolor{white}{#1}};}}

\usepackage{pifont}   % for tick and cross marks
\usepackage{float}    % for table placement
\usepackage{xcolor}   % for colored text
\usepackage[numbers,sort&compress]{natbib}

\usepackage{cleveref}
\crefname{section}{§}{§§}
\Crefname{section}{§}{§§}

\author{%
  \IEEEauthorblockN{Md Hasanur Rashid$^1$, Nathan R. Tallent$^2$, Forrest Sheng Bao$^3$, Dong Dai$^1$}
  \IEEEauthorblockA{$^1$University of Delaware, $^2$Pacific Northwest National Laboratory, $^3$Iowa State University\\
  Email: \{mrashid,dai\}@udel.edu, tallent@pnnl.gov, forrest.bao@gmail.com
  }
}

\begin{document}

%\title{Learned Decentralized I/O Auto-tuning via Client-side Local Metrics}
%\title{LightTune: Lightweight I/O Tuning via learned Local Metrics}
%\title{LightTune: Lightweight Application-agnostic Adaptive I/O Tuning}
\title{CARAT: Client-Side Adaptive RPC and Cache Co-Tuning for Parallel File Systems}
%Online I/O Optimization

\thispagestyle{plain}
\pagestyle{plain}
\ifthenelse{\boolean{maketitleAfterAllMeta}}{}{\maketitle}

%Default configurations of PFS often result in suboptimal performance under such variability. 
\begin{abstract}
Tuning parallel file system in High-Performance Computing (HPC) systems remains challenging due to the complex I/O paths, diverse I/O patterns, and dynamic system conditions. 
While existing autotuning frameworks have shown promising results in tuning PFS parameters based on applications' I/O patterns, they lack scalability, adaptivity, and the ability to operate online. 
% In this work, we present \textbf{LightTune}, a lightweight, application-agnostic, and adaptive I/O tuning framework that operates solely at each I/O client based on locally observed metrics.
In this work, focusing on scalable online tuning, we present \texttt{CARAT}, an ML-guided framework to co-tune client-side RPC and caching parameters of PFS, leveraging only locally observable metrics.
% LightTune employs a novel two-stage tuning strategy that dynamically adjusts both RPC and caching parameters based on the I/O activeness of each client.
% Unlike global or pattern-dependent approaches, LightTune enables each client to make independent tuning decisions in real time, ensuring responsiveness to changes of both application I/O behaviors and system states.
{Unlike global or pattern-dependent approaches, \texttt{CARAT} enables each client to make independent and intelligent tuning decisions online, responding to real-time changes in both application I/O behaviors and system states.}
% We implement LightTune on the Lustre and evaluate it across diverse scenarios, including dynamic I/O patterns, real-world HPC workloads, and multi-client deployments.
{We then prototyped \texttt{CARAT} using Lustre and evaluated it extensively across dynamic I/O patterns, real-world HPC workloads, and multi-client deployments.}
The results demonstrated that \texttt{CARAT} can achieve up to \textbf{3$\times$} performance improvement over the default or static configurations, validating the effectiveness and generality of our approach. Due to its scalability and lightweight, we believe \texttt{CARAT} has the potential to be widely deployed into existing PFS and benefit various data-intensive applications.
\end{abstract}
  
\section{Introduction}
\label{sec:intro}

\input{sc/1_intro}

\section{Background and Key Ideas}
\label{sec:background}
\input{sc/2_back}

%\section{Motivation Examples}
%\label{sec:motivation}
%\input{sc/moti}

\section{Design and Implementation} 
\label{sec:design}
\input{sc/3_design}

\section{Evaluation}
\label{sec:eval} 
\input{sc/4_eval}

\section{Related Work}
\label{sec:related}
\input{sc/5_related}

\section{Conclusion and Future Plan}
\label{sec:conclude}
\input{sc/6_conclude}

\section*{Acknowledgments}
We sincerely thank the reviewers for their valuable feedback. We further thank Xinyi Li and Youbiao He for their early contribution in ML models evaluation. This work was supported in part by the National Science Foundation (NSF) under grants CNS-2008265 and CCF-2412345. This effort was also supported in part by the U.S.\@ Department of Energy (DOE) through the Office of Advanced Scientific Computing Research's ``Orchestration for Distributed \& Data-Intensive Scientific Exploration'' and
the ``Decentralized data mesh for autonomous materials synthesis'' AT SCALE LDRD at Pacific Northwest National Laboratory. PNNL %Pacific Northwest National Laboratory
is operated by Battelle for the DOE under Contract DE-AC05-76RL01830. The Authors acknowledge the National Artificial Intelligence Research Resource (NAIRR) Pilot for contributing to this research result. The authors acknowledge the assistance of ChatGPT and Claude (Anthropic) in performing editorial revisions to the manuscript.

\bibliographystyle{unsrtnat}
{
\bibliography{bib}
}

%  \section{Appendix}
%  \label{sec: appendix}
%  \input{sc/appendix}
\end{document}

%% file: sc/1_intro.tex
\begin{figure}[t]
    \centering
    \includegraphics[width=0.75\columnwidth]{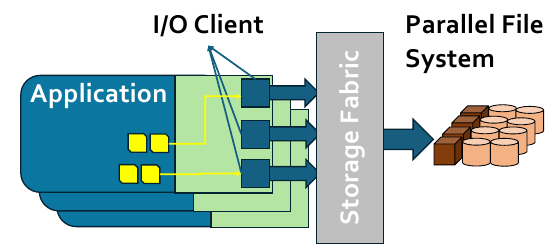}
    \caption{The I/O clients serve individual I/O request.}
    \label{io}
    \vspace{-1em}
\end{figure}

Today, high-performance computing (HPC) systems increasingly support data-intensive scientific applications~\cite{lewis2025machine,parallel__2024,bez2023access,byna2012parallel,clarke20121000}. These applications typically execute across a large number of computing nodes, issuing concurrent I/O requests to multiple remote storage servers. As illustrated in Figure~\ref{io}, the I/O client, which runs within each computing node as part of the parallel file system (PFS) client library, translates application-level I/O requests into individual I/O RPCs over the network, ultimately reaching different storage servers within the PFS. These concurrent I/O clients, operating rigidly according to their predefined parameter settings, can easily cause mismatches among various components along the long I/O paths at runtime, leading to suboptimal performance.
%running concurrently on individual or across computing nodes
For example, an I/O client might attempt to send excessive I/O requests to an already saturated network or storage server causing high contention and degraded performance, or it may fail to construct sufficient I/O RPCs promptly despite available network capacity~\cite{rashid2025adaptbf}.

Essentially, at runtime, the I/O clients must balance two key aspects that significantly impact I/O performance: (1) the rate at which they generate I/O RPCs, and (2) the rate at which these RPCs are transmitted and served. These aspects can be further summarized into two fundamental factors: the \textit{local cache/buffer status} and the \textit{communication performance}.
The state of the local buffer or cache, including its total capacity and current utilization, dictates how application-generated I/O requests are aggregated into locally buffered operations or immediately issued as network I/O RPCs. Meanwhile, the condition of the communication layer, characterized by (1) the number of RPCs that can be sent concurrently, (2) their sizes, and (3) the rate at which remote storage servers process them, determines how efficiently those network I/O RPCs are served. These two factors, like producers and consumers, must operate in coordination to achieve optimal performance.
%After I/Os were formulated from application, the behavior of I/O clients is primarily determined by two factors: 
It has been shown that tuning relevant configurations, such as cache/buffer sizes or the number of concurrent network channels, can significantly enhance the I/O performance~\cite{rashid2023iopathtune,rashid2025dial,egersdoerfer2025stellar}. For instance, Bez et al. demonstrated proper configurations can achieve a \textbf{6.8$\times$} speedup for the \texttt{OpenPMD} application and a \textbf{7.9$\times$} speedup for the \texttt{FLASH-IO} benchmark on the Cori and Summit supercomputers~\cite{bez2021bottleneck}. 
%Xie et al. obtained 7x speedup for \texttt{VPIC-IO} and 3.2x speedup for \texttt{BD-CATS-IO} through cross-layer tuning on Cori~\cite{xie2021battle}.

Most of existing efforts try to find optimal configurations via modeling application I/O patterns and leveraging these patterns to explore configuration spaces. For instance, one may have larger number of RPCs that can be concurrently sent for large sequential I/O patterns because they typically benefit from more parallel network channels, whereas small random I/O patterns may benefit from reducing the maximal size of RPCs, which facilitates smaller RPCs for better network utilization. For instance, Behzad et al. proposed a pattern-driven I/O tuning approach using genetic algorithms guided by I/O pattern prediction models~\cite{behzad2019optimizing}. Other studies have explored different search strategies based on modeled I/O patterns~\cite{agarwal2019active, baugbaba2021improving, liu2023optimizing}. Recent methods, such as CAPES~\cite{li2017capes} and AIOC2~\cite{cheng2021aioc2}, have integrated deep reinforcement learning (DRL) for end-to-end I/O parameter tuning, implicitly modeling I/O patterns through reinforcement learning agents.

However, \textit{understanding application I/O patterns and then configuring I/O clients accordingly} faces practical challenges. HPC applications' I/O patterns arise from the collective behaviors of all ranks. Accurately detecting them is inherently difficult, and effectively using these detected patterns to navigate extensive configuration spaces further complicates the task. The accuracy of I/O pattern detection can be substantially impaired by factors such as changing application I/O behaviors and unpredictable interference from concurrent applications~\cite{egersdoerfer2024understanding,isakov2022taxonomy,egersdoerfer2025ioagent}. Models that demonstrate good performance often rely heavily on global runtime metrics, such as CAPES~\cite{li2017capes} and AIOC~\cite{cheng2021aioc2}, resulting in high overhead for metric collection and model training. Due to such heavy overhead, `best configurations' are frequently simplified to a single set of values to be uniformly applied across all I/O clients, even if they may be from different computing nodes and interact differently with storage servers or network channels~\cite{li2017capes,cheng2021aioc2,zhu2022magpie,lu2022adsts,dong2025rl4sys}. Such simplifications lead to unstable and suboptimal performance, as demonstrated in subsequent evaluation sections.

In this study, we propose a radically different approach. Instead of relying on modeling applications' global I/O patterns, we narrow our focus directly on the individual I/O clients on the computing nodes. Specifically, we tune configurations to alter behaviors of each I/O client purely \textit{\textbf{based on locally observable metrics instead of global application I/O patterns}}. Leveraging machine learning models, multiple independent tunable I/O clients on different computing nodes can make timely decisions to respond effectively to changing workloads and collectively to global storage system conditions, resulting in improved I/O performance for the entire application.

%In this study, we propose a radically different approach. Instead of relying on modeling applications' I/O patterns for tuning, we focus on each I/O client. More specifically, we tune configurations to change the behaviors of individual I/O client purely using the local observable metrics for that I/O client. With the help of machine learning models, multiple tunable units that correspond to different I/O clients of many concurrent applications will receive independent tuning decisions, reacting collectively to what is happening in the global storage systems in a timely manner and achieving better I/O performance for each application.

%The intuition behind is that because the local metrics are collected locally, they directly reflect 1) how the I/O operations are constructed via buffering/caching mechanisms, 2) how they were sent out via network RPCs, and 3) how the global network/storage systems response to the requests. Although such implicit information is complicated to catch manually, if picked and processed correctly, it is learnable. 

The benefits of tuning individual I/O clients are threefold. First, it is extremely lightweight, allowing immediate response to changing I/O patterns or even short-term interferences on the server side. Second, each client is independently tuned, enabling the assignment of optimal configurations per client instead of a single compromised configuration for all clients. Lastly, by treating global network/storage systems as external environments and considering their real-time responses to previous requests, the approach can effectively react to global system status.
%such as runtime contention, cross-server or cross-application interferences, and performance variability, hence enable implicit coordination.
%
We argue that, although the metrics are collected locally, they can implicitly reflect critical information for conducting global tuning. Specifically, they can reflect: 1) how I/O operations are constructed through buffering/caching mechanisms, 2) how these operations are sent as network RPCs, and 3) how global network/storage systems respond to previous RPC requests. Manually interpreting such implicit information is difficult, we can leverage machine learning to learn it.

% To this end, we propose \textcolor{blue}{\texttt{CARAT}}, a novel, lightweight auto-tuning framework designed to dynamically adjust tunable parameters of I/O clients based solely on locally observed client-side metrics. Running on each I/O client, \textcolor{blue}{\texttt{CARAT}} periodically extracts low-level metrics from system stats and employs a trained machine learning model to determine whether better configurations exist to improve I/O performance. If optimal configurations are identified, they are immediately applied. This process repeats at user-defined intervals, ensuring responsiveness to real-time changes in both application I/O behaviors and global storage system conditions.
To validate such an idea, in this study, we designed and prototyped \textbf{C}lient-side \textbf{A}daptive \textbf{R}PC and c\textbf{A}che co-\textbf{T}uning (\texttt{CARAT}) framework for co-tuning \emph{client-side RPC and cache} of Lustre, one of the most popular parallel file systems used in modern HPC clusters~\cite{plechschmidt2020a}. 
%~\cite{braam2019lustre}
Running on each Lustre I/O client, \texttt{CARAT} samples client-local metrics and applies a learned policy to adjust the RPC window, in-flight concurrency, and dirty-cache size online. Tuning decisions are enacted immediately and periodically, enabling rapid adaptation to changing I/O patterns of application and contention cluster-wide.
{We then extensively evaluated \texttt{CARAT} across various scenarios, including sequential/random and dynamic patterns, single- and multi-client runs, and interference conditions,
%targeting data-path client parameters that govern RPC formation and caching. We evaluate across sequential/random and dynamic patterns, single- and multi-client runs, and interference conditions. 
\texttt{CARAT} improves performance by up to \textbf{3$\times$} over default Lustre settings. Our contributions are:}

\begin{comment}
\begin{itemize}[topsep=2pt, partopsep=0pt, itemsep=2pt, parsep=0pt]
    \item We developed a novel autotuning framework that is capable of online adaptive parameter tuning independently and relies solely on local metrics.
    \item We identified key learned low-level metrics derived from clients that effectively characterize the status of clients' I/O activity and the global storage systems.
    \item We conducted comprehensive evaluations across diverse scenarios and demonstrated performance improvement up to 6x over baseline execution. 
\end{itemize}
\end{comment}

\begin{itemize}[topsep=2pt, partopsep=0pt, itemsep=2pt, parsep=0pt]
%mechanism-grounded? @dong what is this
    \item A ML-guided system for \emph{client-side RPC and cache co-tuning} leveraging client-local observations.
    \item A client-side observability model that enables online, per-client decisions without cluster-wide instrumentation.
    \item Comprehensive evaluations demonstrating up to \textbf{3$\times$} improvement over default configuration across diverse scenarios.
\end{itemize}

The rest of the paper is organized as follows: In \S\ref{sec:background}, we discuss relevant backgrounds of Lustre file system and key ideas behind some of \texttt{CARAT}'s design choices. In \S\ref{sec:design}, we present the architecture of \texttt{CARAT} in detail. We present the extensive experimental results in \S\ref{sec:eval}. We highlight the relevant works to our study in \S\ref{sec:related}, conclude the paper and discuss the future work in \S\ref{sec:conclude}.

%% file: sc/2_back.tex
% To demonstrate the \textit{feasibility} and \textit{effectiveness} of \texttt{LightTune}, we conducted evaluations using Lustre, one of the most widely used parallel file systems in practice~\cite{braam2019lustre}.
%In this section, we first describe Lustre’s architecture and operation, highlighting how our proposed approach can be implemented to configure Lustre I/O clients. Building upon this foundation, we introduce the core ideas underlying \texttt{CARAT}. 
%We also discuss how our approach can be generalized and applied to other PFSs, such as NFS, later in this section. 

%{Lustre~\cite{braam2019lustre} is one of the most widely used parallel file systems among top HPC facilities worldwide~\cite{plechschmidt2020a}. This section proceeds as follows: we first review Lustre’s architecture and the client I/O path with RPC formation; we then define the client-side tunables we control, analyze RPC and caching bottlenecks, and motivate \texttt{CARAT} via client-local metrics and an ML-based control strategy; we conclude with applicability to other PFSs.}
In this section, we first use Lustre~\cite{plechschmidt2020a} as an example to illustrate how the local cache and network RPC mechanisms interact and jointly influence I/O performance. We then discuss how local metrics and machine learning models can be leveraged to capture such complex interactions. Finally, we briefly discuss the applicability of our approach to other parallel file systems.

\subsection{Lustre RPC and Caching Mechanisms}

\begin{figure}[t]
    \centering
    \includegraphics[width=\columnwidth]{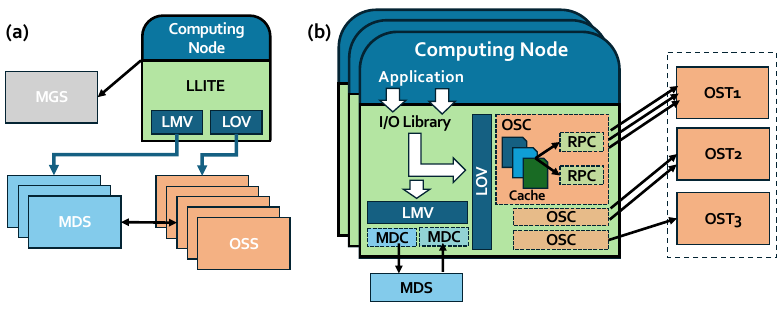}
    \caption{(a) The overall architecture and (b) the detailed I/O path of Lustre.}
    \label{lustre-overall}
    \vspace{-1em}
\end{figure}

Figure~\ref{lustre-overall}(a) first illustrates the overall architecture of Lustre which comprises one management server (MGS), one or more metadata servers (MDSs), and multiple object storage servers (OSSs). The management server (MGS) is lightweight and typically deployed alongside one of the metadata servers (MDSs). On a computing node, I/O requests issued by applications, such as \texttt{open} or \texttt{read}, are managed by the I/O Client library. Within this library, Lustre Lite (LLITE) acts as a bridge between application and the underlying Lustre infrastructure, represented by the Lustre Object Volume (LOV) and the Lustre Metadata Volume (LMV). 
%Lustre utilizes remote procedure calls (RPCs) to facilitate communication between clients and servers.
%{\color{red} figure3/4 needs updates}.

% \subsection{Lustre I/O Clients}
%\textcolor{blue}{\subsection{Lustre I/O Path and RPC Formation}}
Figure~\ref{lustre-overall}(b) further shows how I/O requests are processed by the I/O clients. For data accesses, the I/O requests are handled by the LOV component, which establishes multiple Object Storage Client (OSC) interfaces, each corresponding to one Object Storage Target (OST) on OSS. Each OSC interface is responsible for formulating RPCs and managing RPC communications directed to a specific OST. Similarly, the Lustre Metadata Volume (LMV) component manages metadata accesses to metadata targets (MDTs) on MDSs via the Metadata Client (MDC) interface. Together, OSC and MDC play the role of \textit{I/O Client} as described earlier.
%Note that, in the following sections, we will use \textbf{\texttt{Data I/O Client}} and \textbf{\texttt{Metadata I/O Client}} to name OSC and MDC respectively. 
%{Note that, in the following sections, we will use \textbf{\texttt{Data I/O Client}} to refer to OSC.}

%The interface that translates the application's I/O requests to the corresponding OSC interfaces is called LLITE. 
% To enhance I/O performance, the Data I/O Client maintains a local buffer to cache recently written (`dirty') data. These individual write requests will be temporarily buffered and aggregated into 

During I/O operations, the I/O Client will utilize local \emph{dirty-page cache} to buffer individual writes, which will be aggregated into RPCs to be sent to the remote OSTs.
%For read requests, a readahead mechanism at the LLITE layer proactively prefetches data to optimize performance.
% The actual RPC requests sent to the remote storage servers are constructed by the Data I/O Client after the buffering and caching stages.
%{After buffering, the client constructs the RPCs that will }
% Specifically, RPCs are built based on a fixed-size extent, defined as `\textit{RPC Window Size}', representing a consecutive set of pages targeting the same storage target (OST).
% \textcolor{blue}{RPCs are built from a fixed-size \emph{RPC extent} (the \textit{RPC window size}): a consecutive group of pages targeting the same OST. Each extent maps to a contiguous block of data on the server, and all bytes within an extent are combined into a single RPC.}
RPCs are built based on building a fixed-size \emph{RPC extent} (size controlled by \texttt{osc.*.max\_pages\_per\_rpc}), which includes a consecutive group of pages targeting the same OST. 
%Each extent maps to a contiguous block of object data on the server.
%, and all bytes within an extent are combined into a single RPC.
% I/O requests' data within the same extent will be combined into a single RPC.
% Once RPCs reaches the predefined size or has been waiting beyond a certain threshold, the Data I/O Client will dispatch it to the corresponding storage targets (OSTs).
{RPCs will be mature and dispatched via network channels when one of following events occur: 1) the extent fills, 2) a kernel wait threshold expires, or 3) cache pressure (\emph{cache-waiters}) forces a flush. How many RPCs can be sent out concurrently is bounded by \textit{RPCs in flight} (controlled by \texttt{osc.*.max\_rpcs\_in\_flight}).

{Figure~\ref{fig:rpc-formation}(a) details how the ideal case works. It shows a 64\,kB request issued at a 64\,kB offset gets split into 4\,kB pages and packed into 16\,kB RPC extents. In this example, because the request is sequential and aligned to page and extent boundaries, all extents are filled and triggering the resulting RPCs to be issued as soon as an RPC channel is available.}
\begin{figure}[t]
    \centering
\includegraphics[width=\columnwidth]{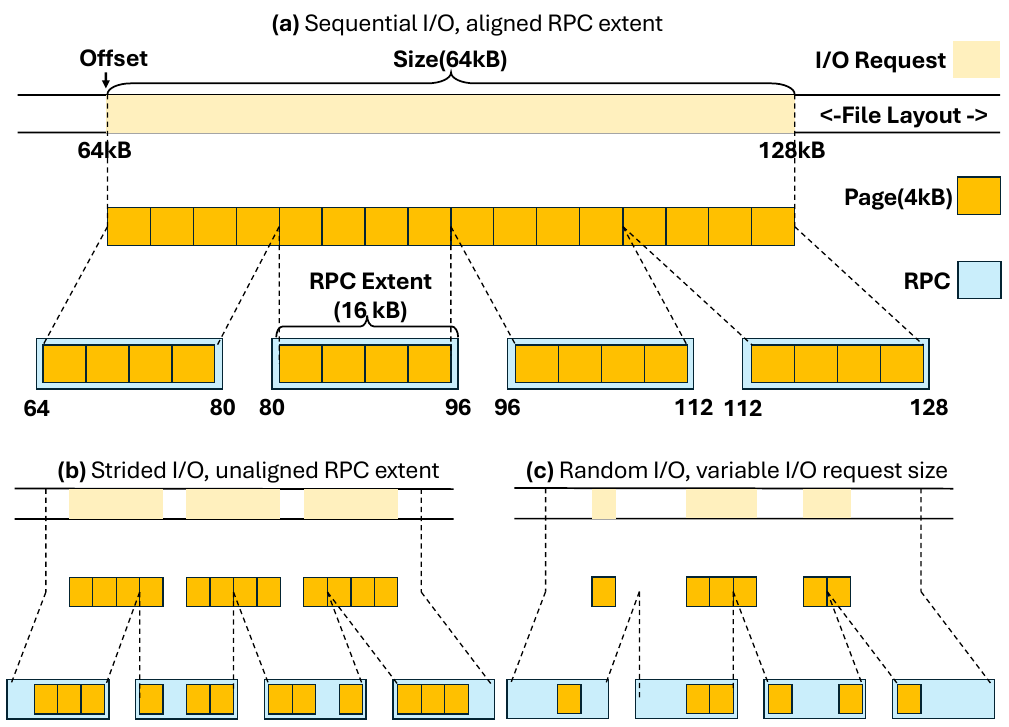}
    \caption{{An illustration of how I/O requests are split into pages and aggregated into fixed-size RPC extents.}}
    \label{fig:rpc-formation}
    \vspace{-1em}
\end{figure}

%\textcolor{blue}{\subsection{RPC and Caching Bottlenecks}}

% \textcolor{blue}{Lustre’s client-side buffering and fixed-extent RPC formation introduce recurrent bottlenecks that degrade performance under diverse I/O patterns and dynamic contention. These issues arise from fundamental mechanism-level behaviors rather than misconfiguration.}
\textbf{Complex coordination between RPC and Caching.} {In practice, the buffering and RPC formation could introduce various I/O performance bottlenecks under diverse I/O patterns and dynamic contention as exemplified below.} %These issues arise from fundamental mechanism-level behaviors.

\textit{a) Under-filled extents and cache fragmentation.}
Irregular, unaligned, or random I/O requests may not be able fill RPC extents in time. These partially filled extents will be held in cache, occupying space, blocking admission of new pages, and delaying sending out I/Os. 
%I/O is acknowledged once cached, but RPC dispatch is deferred until the extent fills, a kernel timeout expires, or cache-waiters force a flush. 
\cref{fig:rpc-formation}(b) and (c) show examples of such cases, where unaligned strided or random small I/Os are organized into different extents. These partially filled extents will not be sent out as a new RPC until a timeout event happens, which delays the existing I/Os while blocking the new I/Os. %how extent-partial sequential requests (a), unaligned offsets (b), striding (c), and small/random accesses (d) increase the share of under-filled RPCs compared to aligned sequential I/O (Figure~\ref{fig:rpc-formation}).}

% \paragraph{Premature small-RPC dispatch}
% \textcolor{blue}{When the RPC window is too small or dispatch thresholds are aggressive, many RPCs are sent before reaching full size. This produces a surge of small RPCs, each paying fixed per-RPC network and server-queue costs and reducing coalescing opportunities for subsequent I/O.}

\textit{b) Server-side congestion under contention.}
{On the other hand, if I/O Clients are able to fill the extents fast and construct RPCs radically, these bursts of small RPCs can overwhelm the remote storage server's I/O queues, significantly increasing queuing delay and degrading throughput. Worse, the congestion could further amplify the impacts of premature dispatch and insufficient extent filling.}

\textit{c) Cache-limit throttling.}
{The size of the dirty caches also impacts the I/O behaviors. The conservative, low dirty-cache limits will send out immature RPCs earlier, which restrict coalescing and cap throughput. But, overly large dirty-cache limits can buffer lots of I/Os, leading to bursty flush storms that overwhelm the backend during transient write spikes.}

As these examples show, the interactions between caching and RPC are complicated and impacted by various factors, such as workloads, server status, and network contentions. In practice, there may be additional RPC classes (e.g., cache-waiters or lock-conflict related) further complicating the coordination to optimize I/O performance. In this study, we propose to leverage machine learning methods to learn such complex relationships. Model details are in \S~\ref{subsec:ML_model}

\begin{comment}
\begin{figure}[t]
    \centering
\includegraphics[width=0.4\textwidth]{sc/figs/rpc-underutilization-scenarios.png}
    \caption{{Extent under-utilization across access patterns—misalignment, striding, and small/random requests fragment the cache and delay RPC formation.}}
    \label{fig:rpc-underutilization}
\end{figure}
\end{comment}

% \subsection{Lustre Parameters for Autotuning}
{\subsection{Tunable Parameters for RPC and Caching}}
% Based on how Lustre I/O Clients handle I/O requests, we selected a set of tunable parameters to validate our idea in this study. These parameters belong to two categories: \textit{RPC Parameters} and \textit{Caching Parameter}. The Table~\ref{tab:lustre_param_summary} summarizes the tunable parameters.
%Based on how Lustre I/O Clients handle I/O requests and how Caching and RPC interacts, 
In this study, we focus on two categories of parameters to tune, as Table~\ref{tab:lustre_param_summary} summarizes. {Note that, while metadata requests follow similar RPC and cache structure, in this work we focus on data-path behavior and do not tune metadata parameters.}

\setlength{\belowdisplayskip}{-10pt}
\begin{table}[t]
\footnotesize
\centering
\caption{Summary of Tunable Parameters for Lustre Clients}
\begin{tabularx}{\columnwidth}{l l X}
\toprule
\textbf{Category} & \textbf{Parameter} & \textbf{Description} \\
\midrule
\multirow{2}{*}{RPC}
  & \texttt{osc.*.max\_pages\_per\_rpc} & Max size of a single data RPC. \\
  & \texttt{osc.*.max\_rpcs\_in\_flight} & Max number of concurrent data RPCs allowed. \\
\midrule
Caching  & \texttt{osc.*.max\_dirty\_mb} & Max dirty-page cache limit for data I/O. \\
\bottomrule
\end{tabularx}
\label{tab:lustre_param_summary}
\vspace{-1em}
\end{table}

\begin{comment}
\setlength{\belowdisplayskip}{-10pt}
\begin{table}[htpb]
\footnotesize
\centering
\caption{Summary of Tunable Parameters for Lustre Clients}
\begin{tabularx}{0.48\textwidth}{l l X}
\toprule
\textbf{Category} & \textbf{Parameter} & \textbf{Description} \\
\midrule
\multirow{4}{*}{RPC}
  & \texttt{osc.*.max\_pages\_per\_rpc} & Max size of a single data RPC. \\
  & \texttt{osc.*.max\_rpcs\_in\_flight} & Max number of concurrent data RPCs allowed. \\
  & \texttt{mdc.*.max\_pages\_per\_rpc} & Max size of a single metadata RPC. \\
  & \texttt{mdc.*.max\_rpcs\_in\_flight} & Max number of concurrent metadata RPCs allowed. \\
\midrule
Dirty Page  & \texttt{osc.*.max\_dirty\_mb} & Max cache limit for data I/O. \\
Cache  & \texttt{mdc.*.max\_dirty\_mb} & Max cache   limit for metadata I/O. \\
\bottomrule
\end{tabularx}
\label{tab:lustre_param_summary}
\end{table}
\end{comment}

We selected these parameters for several reasons. First, they are specific to the I/O clients, allowing them to be tuned independently at each I/O client (\emph{independent}). Second, these parameters can be dynamically adjusted and will take affect at runtime, making them ideal for validating our approach (\emph{adaptive}). In contrast, many Lustre parameters, such as \textit{stripe count} and \textit{stripe size}, cannot be modified once set, and thus are excluded from our considerations. Our tuning can complement these tuning efforts at runtime. For instance, 
\texttt{CARAT} can further enhance I/O performance after the optimal stripe count and stripe size have been determined. 
%dong: is this repeated to Second?
%Finally, all the selected parameters take effect at runtime (\emph{online}), enabling % \texttt{LightTune} \texttt{CARAT} to respond immediately to workload changes and runtime interference.

It is also important to highlight that, although both RPC parameters and caching parameters take effect at runtime, their impacts manifest at different rates. 
Changes to RPC parameters affect I/O operations immediately, influencing all newly constructed RPCs. Although existing RPCs already in flight remain unaffected, they typically complete within microseconds; therefore the delay is negligible. 
%On the other hand, changes to caching parameters require more time to propagate. For instance, increasing buffer size may not show performance changes until existing buffered RPCs being sent out, a process that could take seconds. 
On the other hand, changes to caching parameters require more time to propagate. For instance, increasing cache size may not show performance changes until sufficient I/O arrives to take advantage of increased cache size.
This discrepancy makes it challenging to determine whether performance changes are due to adjustments in RPC or caching parameters. To address this, \texttt{CARAT} employs a two-phase tuning approach, handling these parameters separately. The details will be discussed in \S~\ref{subsec:two-stage}.

\subsection{Applicability to Other PFS}
% \texttt{LightTune} works for any PFS as long as it contains three features: client-server I/O communication, dynamically tunable parameters within the I/O Clients, and observable local metrics. These features widely exist in various HPC storage systems. 
Our approach targets a rather general design principle in parallel file systems: buffered, aggregated small I/Os served via network RPCs. Such a practice is widely seen and essential for better I/O performance in networked FS. Hence, we argue the method can be applied widely.
%generalizes to systems that share three properties: client–server I/O over RPC, dynamically adjustable client-side parameters, and observable local metrics.}
% For instance, the Network File System (NFS)~\cite{shepler2003network}, another popular file system in HPC, leverages RPCs to conduct I/Os between clients and servers. 
{For example, Network File System (NFS)~\cite{shepler2003network} buffers and issues I/O via RPCs and exposes analogous tunables.}
% NFS also offers similar tunables to control the I/O clients. For instance, `\texttt{nfs4\_max\_transfer\_size}' determines the maximum data transfer size and `\texttt{tcp\_slot\_table\_entries}' sets the maximum number of RPC requests, both of which have effects similar to their counterparts in Lustre. 
{Parameters such as \texttt{nfs4\_max\_transfer\_size} (max transfer size) and \texttt{tcp\_slot\_table\_entries} (max outstanding RPCs) play roles similar to \texttt{max\_pages\_per\_rpc} and \texttt{max\_rpcs\_in\_flight}.}
% NFS tools like `\textit{nfsiostat}' allow tracking performance-related metrics. These local metrics and tunable parameters allow our proposed idea to be extended to NFS.
{Client-local tools (e.g., \textit{nfsiostat}) provide the metrics needed for online decisions, suggesting our mechanism can be adapted where similar client-side RPC and caching behaviors exist.}

%% file: sc/3_design.tex
%To summarize, tuning RPC parameters allows us to handle dynamic changes in I/O patterns, such as bursts, by increasing RPC issuance. Meanwhile, tuning caching parameters enables better absorption of application I/O bursts and supports higher volumes of RPCs simultaneously in-flight. Additionally, the dirty page cache temporarily stages pages, reducing underutilized RPC transmissions, particularly with random access patterns. However, naively increasing cache limits risks overloading servers due to large write flushes during high-concurrency periods. Therefore, large HPC systems restrict cache sizes because the number of interfaces (OSCs/MDCs) can be in the hundreds to thousands. By dynamically tuning caching parameters, we allocate larger cache limits to active interfaces and reduce limits for inactive ones, enhancing the effectiveness of RPC parameter tuning in responding to varying application I/O demands.

%[Dai]. These are great points. I revise the texts to reflect why we tune caching parameters this way. Autocompute is a great idea. We should consider it in the next version.
%\textcolor{orange}{[NRT] 1. Agree with prior comment.
%2. Why would Cache tuning be relavent when I/O is inactive?
%3. Why will this design work well for different applications?
%4. Since you can detect active and inactive, could you autocompute the tuning interval based on how much time it takes to amortize adjustment of RPC? Then you could really claim autonomous and application-agnostic.}

\begin{figure}[t]
    \centering
\includegraphics[width=\columnwidth]{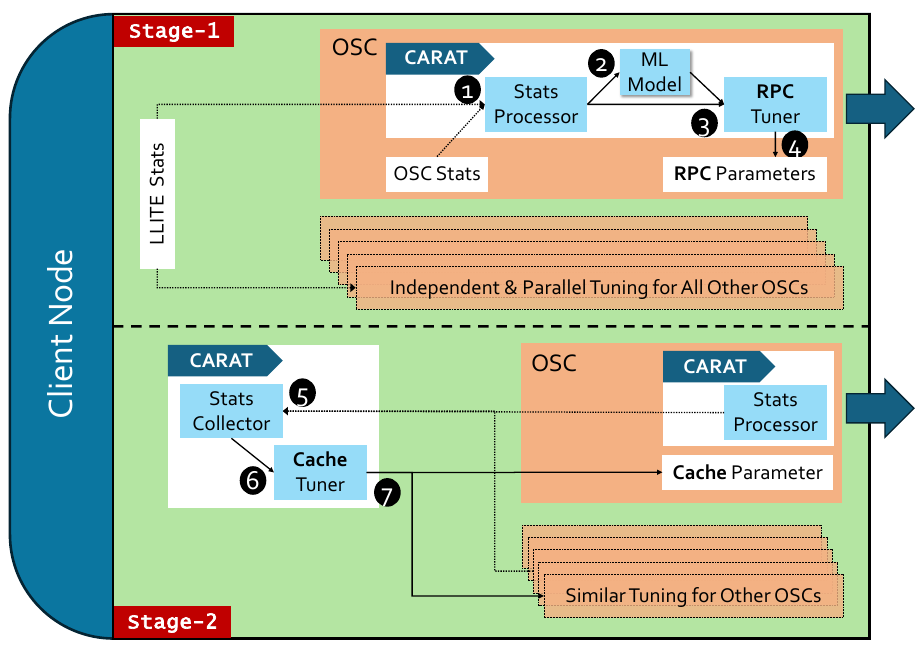}
    \caption{Architecture of the \texttt{CARAT} framework.}
    \label{fig:iotuning}
    \vspace{-1em}
\end{figure}

In this study, we implemented a \texttt{\texttt{CARAT}} prototype based on the Lustre file system, which operates autonomously as part of the Lustre I/O client. 
Figure~\ref{fig:iotuning} illustrates the overall architecture of \texttt{CARAT}, divided into two stages. 

The \textit{Stage-1} begins from the system stats processor component (\circled{1}) probes the local-level statistics from I/O client at user-defined intervals. These statistics are then preprocessed to extract defined metrics (defined in \S~\ref{subsec:metrics}). We then create the snapshots incorporating these metrics for each I/O client. The processed snapshots (\circled{2}) are then fed into the local ML model component, which generates a probability distribution across the configuration space. This distribution assesses the likelihood of each configuration achieving performance improvement if applied. Armed with the probability distribution and the snapshot for each I/O client, the RPC tuner (\circled{3}) component determines the optimal RPC configuration and applies it (\circled{4}). This whole process repeats at each interval and continues throughout the duration of \texttt{CARAT} operation on the client, ensuring continuous optimization.

The \textit{Stage-2} happens concurrently with \textit{Stage-1} at the beginning of I/O active stage following the I/O inactive stage. It focuses on tuning the cache parameters. The stats collector component (\circled{5}) gathers statistics from the stats processor of each I/O client. It then forwards the collected data to the cache tuner component (\circled{6}), which determines the appropriate cache limit for each I/O client based on the observed metrics and applies the corresponding parameter changes (\circled{7}).

\subsection{Two-Stage Tuning Principle}
\label{subsec:two-stage}

\begin{figure}[t]
    \centering
\includegraphics[width=0.75\columnwidth]{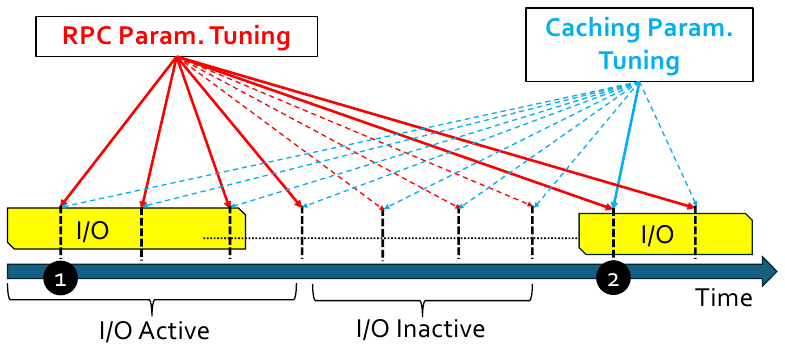}
    \caption{The Two-stage tuning strategy in \texttt{CARAT}.}
    \label{fig:2phase}
    \vspace{-1em}
\end{figure}

\texttt{CARAT} introduces this two-stage tuning strategy to tune RPC parameters and caching parameters separately. 
Figure~\ref{fig:2phase} shows how the two-stage tuning strategy works in \texttt{CARAT}. We consider each application execution would consist of two stages: \textit{I/O Active} and \textit{I/O Inactive}, which can be easily differentiated from the I/O client by observing how many I/O requests were sent to the I/O library during last observation window (e.g., 1 second). If there are I/O requests (yellow part), then we consider to be in the I/O active stage. Otherwise, we define it as the I/O inactive stage. 

During I/O active stage, we only perform RPC tuning based on the user-defined tuning frequency. Since changes to RPC parameters impact the I/O performance instantly, we will be able to observe how previous tuning impacts the system by monitoring the metrics in real-time. Using the most recent local metrics, the tuner makes new tuning decision at each observation (solid arrows, marked \circled{1} in Figure~\ref{fig:2phase}). During I/O inactive stage, there is no actual I/O. We disable RPC parameter tuning during this stage since there exists no RPC transfers.

We tune caching parameter only at the end of the I/O inactive stage, that is, at the first observed time-point when I/O activity resumes (solid arrow, marked as \circled{2} in Figure~\ref{fig:2phase}). 
At this point, the tuner uses collected metrics during \textit{I/O active stage} to assess whether the previous caching configuration is appropriate and adjusts them as needed. 
We select this time point to tune caching parameters because after the period of I/O inactive stage (>1 second), the impacts of the previous caching configuration should have well faded, making the impacts of the new caching configuration more easily observed.

\subsection{I/O Client Local Metrics}
\label{subsec:metrics}

\begin{table*}[t]
\footnotesize
\centering
\caption{Summary of Designed Client-Side Local Metrics and Their Characteristics}
\begin{tabularx}{\textwidth}{XXXXXXX}
\toprule
\textbf{Metric Name} & \textbf{RPC Page} & \textbf{RPC Channel} & \textbf{Unit Page} & \textbf{Data Transfer} & \textbf{Dirty Cache} & \textbf{Estimated Cache} \\
& \textbf{Utilization} & \textbf{Utilization} & \textbf{RPC Latency} & \textbf{Volume} & \textbf{Utilization} & \textbf{Update} \\
\midrule
\textit{Used For:} &
RPC \& Cache tuning &
RPC \& Cache tuning &
RPC tuning &
RPC \& Cache tuning &
RPC \& Cache tuning &
RPC tuning \\
\midrule
\textit{Meaning:} &
Ratio of average to max pages per RPC &
Ratio of average to max RPCs in flight &
Avg. latency per page for RPCs &
Total data transferred through RPCs &
Dirty cache usage ratio &
Estimated data update in cache \\
\midrule
\textit{Captures Locally:} &
RPC formulation quality &
RPC parallelism, actual vs. allowed RPCs &
- &
Volume of transferred data &
Current buffer utilization &
Local cache updates, reduced RPC needs \\
\midrule
\textit{Reflects Globally:} &
- &
Network contention, server grants for RPCs &
Network latency, impact of server I/O load &
End-to-end throughput, total RPC traffic &
Limited allocation due to server overload &
- \\
\bottomrule
\end{tabularx}
\label{table:metrics_summary}
\vspace{-1em}
\end{table*}

% Obtaining low-level client-side metrics is crucial for \texttt{CARAT}: they must expose application I/O behavior and the client’s view of system state with minimal overhead.
Obtaining low-level client-side metrics is crucial for \texttt{CARAT}. The metrics must (i) expose application I/O behavior and the client’s view of network/server state, (ii) scale with negligible overhead, and (iii) be robust enough for online learning and control.
%
% Unlike prior pattern-centric approaches~\cite{li2017capes, cheng2021aioc2, zhu2022magpie}, we (i) use scalable, client-local signals that still reveal application and system effects (\S~\ref{sec:eval}), and (ii) pre-process raw stats into stable features suitable for learning.
Unlike prior pattern-centric approaches~\cite{li2017capes, cheng2021aioc2, zhu2022magpie}, we rely on scalable, client-local metrics that reflect application and system effects (\S~\ref{sec:eval}); we also pre-process raw counters into stable, normalized features suitable for learning.
% All metrics are computed separately for \texttt{read} and \texttt{write} due to differing paths (e.g., dirty-page vs. readahead). Table~\ref{table:metrics_summary} summarizes the set.
All metrics are computed separately for \texttt{read} and \texttt{write} because the paths differ (e.g., dirty-page vs.\ readahead). Table~\ref{table:metrics_summary} summarizes the set; we describe each briefly below.

\textbf{RPC Page Utilization.}
% Ratio of average to maximum pages per RPC; indicates extent fill quality under spatial/temporal access patterns, cache availability, and striping.
This is the ratio of average to maximum pages per RPC. It reflects how well fixed-size extents are filled before dispatch, integrating spatial/temporal access patterns (sequential/strided/random, bursty vs.\ steady), available client cache, and object striping. Persistently low values indicate under-filled extents (more small RPCs, delayed coalescing), whereas high values suggest effective aggregation.

\textbf{RPC Channel Utilization.}
% Ratio of average to maximum \emph{RPCs in flight}; captures realized parallelism vs.\ limits and priority exceptions (e.g., cache-waiters/lock-related RPCs) that may exceed the nominal cap.
Defined as the ratio of average to maximum \emph{RPCs in flight}, this metric captures realized parallelism relative to configured limits. It decreases when RPC formation stalls (e.g., sparse arrivals, cache pressure) or when server-side throttling reduces concurrency; it may exceed nominal caps due to priority issuance (e.g., cache-waiters or lock-related RPCs), signaling exceptional drain behavior.

\textbf{Unit Page RPC Latency.}
% Average per-page RPC latency; normalizes out RPC size and concurrency so variations reflect network/server conditions rather than batching.
Average latency per page transferred, which normalizes out batch size and concurrency. By factoring out RPC size and channels, this metrics reflects network and server conditions (e.g., queue buildup, transient congestion) rather than batching artifacts, thus more stable across pattern shifts.

\textbf{Data Transfer Volume.}
% Bytes transferred over RPCs per interval; provides a direct throughput signal to stabilize decisions.
Total bytes transferred via RPCs during the interval (e.g., 1\,s). It provides a direct throughput signal and, together with latency/utilization, distinguishes “busy and healthy” from “busy and congested” conditions, stabilizing policy updates.

\textbf{Dirty Cache Utilization.}
% Fraction of dirty-page cache in use; indicates pressure/backlog and, with other signals, whether cache headroom is constraining coalescing.
Fraction of the dirty-page cache currently in use. Rising pressure indicates backlog (RPC issuance not draining writes fast enough) or a phase with heavy write arrival; in conjunction with page/channel utilization it reveals whether limited headroom is preventing effective coalescing.

\textbf{Estimated Cache Update.}
% Estimate of bytes updated in the dirty cache (e.g., in-place/small random writes) rather than sent immediately; approximated from per-client write shares to avoid per-request tracing.
An estimate of bytes updated in-place within the dirty cache (e.g., in-place or small random writes that overwrite cached regions) rather than sent immediately. We approximate it from per-client write shares on a node minus observed RPC drain and cache delta, avoiding per-request tracing while still identifying phases where enlarging cache headroom materially improves performance.

\textbf{Metrics on Changes.}
% We also track deltas across intervals to capture short-term trends, enabling responsive, trend-aware decisions.
For all metrics, \texttt{CARAT} also tracks short-term deltas between consecutive intervals. These trends highlight emerging contention or recovery and let the controller react promptly without overcorrecting to transient spikes.

%Our novel designed metrics capture robust information and reveal critical bottlenecks relevant to RPC communication between servers and clients, including considerations for the state of the local buffer. While several previous studies~\cite{li2017capes, cheng2021aioc2, zhu2022magpie} have collected raw statistical information pertinent to RPC, they did not extend their research to further refine the raw statistics. The amalgamation of the designed metrics enables the machine learning model to gain a more concrete understanding of the system's behavior. This enhanced comprehension allows the model to pinpoint bottlenecks more accurately and guide the tuning process more effectively, thereby optimizing the interaction between servers and clients in the HPC environment.

\subsection{ML Model for RPC Parameter Tuning}
\label{subsec:ML_model}
%{\color{red} should we include a figure about how neural network is structured, how the inputs are given, how they run in the client side library, what kind output it produces, and how it can be cooperatred?}
To accurately capture the characteristics of client-local metrics, we utilize ML models to comprehend {how the current values of client metrics might influence I/O and how a particular configuration would impact the performance accordingly}. 

Rather than naively forecasting the exact performance value for a configuration, which is highly dependent on the cluster hardware, we re-map the problem to a classification task. Specifically, the ML model will only determine whether a given configuration will lead to a better I/O performance than the current configuration. We consider `better' as an improvement beyond a set threshold (in our case, 15\%) considering possible noises. 
If the model predicts a greater than 80\% probability of improvements, it classifies the outcome for that configuration as positive (one); otherwise, it is negative (zero). 
The probability threshold defined here (80\%) is to ensure we have a good set of candidates with a high chance of better performance and avoid selecting opportunistic choices with low confidence in performance improvement. %This allows the tuning strategy to confidently identify an optimized configuration candidate, which will be discussed later.

To formally define the ML models, let $\mathbf{s}_t$ represent a vector of client-local metrics at timestep $t$. A short history of system performance of $k$ timesteps is thus $\mathbf{H}_t= [\mathbf{s}_{t-k}, \mathbf{s}_{t-k+1}, \cdots, \mathbf{s}_t]$. We further denote the tunable parameters of the system at timestep $t$ as $\mathbf{\theta}_t$, which is also a  vector. The goal of a ML model here is to find a function $f$ that predicts the likelihood that the setting $\theta_t$ will result in a performance improvement over a threshold $\epsilon$ (e.g., 15\%) in the next timestep, thus $f(\theta_t, \mathbf{H}_t) = \mathbb{E}[ s_{t+1}/ s_{t} > 1+\epsilon]$ where $\mathbb{E}$ is expectation. 

In our study, we experimented multiple ML architectures, including support vector machines (SVMs)~\cite{cortes1995support}, neural networks (NNs)~\cite{mcculloch1943logical}, and gradient boosting decision trees (GBDTs)~\cite{friedman2001greedy}. For NNs, we experimented, Fully-Connected NNs (FC-NNs), vanilla Recurrent Neural Networks (RNNs)~\cite{rumelhart1986learning}, and Temporal Convolutional Networks (TCNs)~\cite{bai2018empirical}. Because SVMs, GBDTs, and FC-NNs do not have the ability to take data sequentially or iteratively, $\mathbf{H}_t$ is flattened into a 1-D vector $s_{t-k} \oplus s_{t-k+1} \oplus \cdots \oplus s_{t}$ before being fed into the said ML architectures. For RNNs and TCNs, the vector $s_{i\in [t-k..t]}$ is fed into the NN each time. As to be shown later in \S~\ref{sec:ML_eval},  GBDTs are the most suitable architecture for our problem and $k=1$ provides the best value.

% The ML model is trained with a set of data samples collected when running benchmark using constantly changing I/O Client configurations. Each time the parameter changes, we check the I/O performance and label that change (as one data sample) as one or zero.
% Once the models is trained, it will work in real time for inference. Each time being queried, the model generates a probability distribution for a given configuration set and the observed low-level metrics. The parameter tuning algorithm then takes this probability distribution, alongside observed low-level metrics, to identify the best configuration to apply for enhancing I/O performance.
We train the model by sweeping client configurations; after each change, we measure the next interval’s performance and label the sample \(\,1\,\) if improvement \(>\!15\%\), else \(0\). At runtime, each client periodically queries the model with \(\mathbf{H}_t\) and candidate \(\boldsymbol{\theta}_t\); the model returns the probability of being 1.

We train separate models for read and write operations. This differentiation stems from observing distinct approaches the Lustre employs in handling these operations. 
%Write operations are subject to intricate RPC formation rules influenced by factors such as the distributed locking mechanism, grant utilization, and the OSC's cache utilization constraints, which are more complex than those for read operations. We design metrics specific to the operation type to capture operation-specific characteristics more precisely. 
By training separate models, we can better predict the likelihood of I/O performance improvement based on client-local metrics and given configuration. In the next section, we discuss how the observation of client-local metrics and insights generated from the ML model enable us to decide the tuning actions on the parameters.

\begingroup
\setlength{\textfloatsep}{0pt}
\begin{algorithm}[t]
\caption{RPC Parameter Tuning Algorithm}
\label{alg:calc_score}

\KwData{RPC space $\Theta$, system history $\mathbf{H}_t$, op type $o$, current config $\theta_t{=}\{\theta_t^1,\theta_t^2\}$, threshold $\tau$, weights $\alpha,\beta$}
\KwResult{Optimized config $\theta^*$}

$S \gets \{\theta \in \Theta\ |\ f(\theta,\mathbf{H}_t) > \tau\}$\;
Normalize $S$ using MinMax normalization\;

$\theta^* \gets \arg\max_{\theta \in S} \big( o = \text{`Write'} \ ?\ \texttt{WriteScore}(\theta, \mathbf{H}_t) : \texttt{ReadScore}(\theta, \mathbf{H}_t) \big)$\;

\SetKwFunction{FWriteScore}{WriteScore}
\SetKwProg{Fn}{Function}{:}{\KwRet}
\Fn{\FWriteScore{$\theta, \mathbf{H}_t$}}{
  $f(\theta, \mathbf{H}_t) \cdot (1 + \beta \sum \theta)$\;
}

\SetKwFunction{FReadScore}{ReadScore}
\SetKwProg{Fn}{Function}{:}{\KwRet}
\Fn{\FReadScore{$\theta, \mathbf{H}_t$}}{
  $f(\theta, \mathbf{H}_t)(1 + \alpha \theta^1) + \theta^2$\;
}
\end{algorithm}
\endgroup

\subsection{RPC Tuner: Conditional Score Greedy}
As described earlier, we adopt separate tuning strategies for \texttt{read} and \texttt{write} operations because reads and writes exercise different components of the Lustre I/O path (e.g., buffering/writeback and lock/cache pressure for writes versus request parallelism for reads), leading to different performance sensitivities.
\texttt{CARAT} maintains two trained models (read-focused and write-focused) and selects the model at each observation period based on the dominant observed I/O volume (\texttt{Data Transfer Volume}, in bytes): if read volume dominates, it uses the read model; otherwise it uses the write model.
The ML model assigns a probability to each input configuration, providing an approximation of which configurations are likely to enhance I/O performance. However, identifying the optimal configuration from these high-probability options is nontrivial. At the beginning, we intuitively utilized the \textit{Greedy Tuning} strategy, which picks the parameter configuration predicted with the highest probability to improve the performance. However, we quickly noticed that the high probability does not always lead to the optimal configuration. Instead, it likely picks the safer configuration with little improvement. We then improved the pure greedy strategy to \textit{Epsilon Greedy}, by introducing a small $\epsilon$ value to randomly select the highest probability configuration (exploitation) or pick a random configuration (exploration). The overall performance improves. But it also introduces long delays and high variances in tuning. 
%upper confidence bound, and Thompson sampling. 
In this study, we introduce a \textit{Conditional Score Greedy} approach that (i) filters candidate configurations using the model probability and a threshold $\tau$, and (ii) ranks only the retained candidates using a priority score that combines the probability with a lightweight notion of parameter ``trend'' (i.e., preference for configurations that are more likely to remain beneficial under future variations).
Algorithm~\ref{alg:calc_score} outlines the tuning strategy. 
%In addition, we calculate a score —yielded superior I/O performance improvements in our experiments. Despite the better performance, we observed the framework becoming trapped in local optima, preventing convergence to a near-optimal solution.

%To achieve near-optimal solutions, our tuning strategy incorporates designed metrics, system behavior knowledge, and filtered configurations probability. We established distinct scoring mechanisms for read and write operations to identify the most promising configuration. 
For each configuration $\theta$, the ML model will output the probability of improving the performance by at least 15\%. If the probability is over a preset threshold $\tau$, the corresponding $\theta$ is added to the set $S$ for further consideration (line 1). We set $\tau$ as $0.8$ in our experiment. Selecting the optimal $\theta$ as $\arg\max_\theta f(\theta, \mathbf{H}_t)$ reduces to pure greedy tuning, which prefers high probability, safe configurations. Thus, we add regularization terms to them (lines 5 and 7).
% For candidates in $S$, based on recent workload is write-dominating or read-dominating, we calculate a score (\texttt{WriteScore} or \texttt{ReadScore}). 
% The Algorithm then heuristically picks the configuration with the highest score to apply to the system. 
These regularization terms bias selection toward candidates that are both high-confidence and sufficiently ``progressive'' (rather than minimally safe), which empirically reduces long exploration delays while avoiding unstable random exploration.

Specifically, we prioritize the larger values of the two selected parameters (`\textit{RPC Window Size}' and `\textit{RPCs in Flight}') when several configurations can all improve the performance. This is because, in general, a larger RPC size would utilize network channels better and more RPCs in flight would transfer more data in parallel. Hence, they have a higher chance of being better configurations in the future. Of course, higher values alone certainly would lead to extremely bad performance, which necessitates the continuous tuning. To control this tradeoff, the score uses two hyper-parameters $(\alpha,\beta)$ that balance the importance of the candidate's predicted probability versus the preference term over the $\theta$ values; in our experiments we set $\alpha=\beta=0.5$ to provide a balanced gain--stability tradeoff.
%They are set as $(1, 0.5)$ in our experiment.} 

\begingroup
\setlength{\textfloatsep}{2pt}
\begin{algorithm}[t]
\caption{Cache Parameter Tuning Algorithm}
\label{alg:cache_alloc}

\KwData{Global cache limit $D_{max}$, active I/O clients $O_{act}$, inactive I/O clients $O_{inact}$, write volume $V_w[o]$, max cache usage ${U}_{max}[o]$, max in-flight RPC volume $R_{max}[o]$, discrete values of cache $B=\{2^x\};x{=}5..11$,$B_{\min}=2^{5}$,$B_{\max}=2^{11}$}
\KwResult{Final cache allocations for all I/O clients, $D$}

$b(x):=\min\{b \in B : b \geq x\}$\;
$D[o] \gets B_{\min}$ for all $o \in O_{inact}$\;
$budget \gets D_{max} / |O_{act}|$,\quad $V_{tot} \gets \sum_{o \in O_{act}} V_w[o]$\;

\eIf{$budget \geq B_{\max}$}{
  $D[o] \gets B_{\max}$ for all $o \in O_{act}$\;
}{
  $D[o] \gets \max\left(b({U}_{max}[o]),\ b(R_{max}[o]),\ b(D_{max} \cdot V_w[o] / V_{tot})\right)$ for all $o \in O_{act}$\;
}
\Return{$D$}
\end{algorithm}
\endgroup

\subsection{Cache Tuner: Rule-based Heuristic}
\label{subsec:cache_alloc}
%The Lustre file system buffers write data in a dirty page cache before sending it to servers. Each client-side interface (OSC/MDC) has a dirty cache limit, and Lustre also imposes a global limit, typically half of the client's main memory. Although the per-interface limit usually defaults to two gigabytes, HPC environments commonly have hundreds to thousands of interfaces per client due to numerous OSTs and MDTs. To prevent server overload from large concurrent write flushes, administrators often reduce this limit\textcolor{red}{ref?~\cite{}}, which unfortunately lowers the cache's ability to absorb I/O bursts and increases cache wait latency. However, since not all interfaces remain active simultaneously, setting higher cache limits for active interfaces while reducing them for inactive ones would balance allocation without risking overload. We propose Algorithm~\ref{alg:cache_alloc} to achieve this dynamic allocation.

The Lustre file system buffers write data in a dirty page cache before sending it to servers. As discussed in \S~\ref{subsec:two-stage}, this cache helps absorb I/O bursts and supports higher volumes of RPC transmission. However, naively setting a high cache limit can trigger frequent data flushes and risk server overload during high concurrency across I/O clients, leading production HPC systems to restrict cache sizes. To address this, we propose a dynamic cache limit allocation in Algorithm~\ref{alg:cache_alloc}, which leverages a subset of local metrics collected during RPC tuning to enable informed and adaptive allocation.

%\textcolor{red}{This description is not matching your algorithm well. Need update. For algorithm 2, it is also overly complicated. Need simplifying it. Also, what are the inputs for the rule-based heuristic? What metrics are useful? We should also cover this in the previous and this paragraphs.}
%I have updated both the algorithm and the description to simplify and allow better intuitive understanding. In the Table~\ref{table:metrics_summary}, I have added a row indicating which tuning they are being used for.}
The algorithm begins by assigning the lowest cache value to all I/O clients with no I/O activity (line 2). If the available cache exceeds the total that active I/O clients can maximally use, it assigns the maximum cache value to all active I/O clients (line 5). Otherwise, it considers three factors to determine each active I/O client’s cache allocation: (1) \textit{maximum cache utilization}, reflecting I/O bursts absorbed by the cache; (2) \textit{peak in-flight RPC volume}, indicating RPC bursts accommodated; and (3) \textit{the proportion of write RPCs handled by the I/O client}. For each factor, the algorithm selects the nearest equal or higher discrete cache value and then assigns the maximum among these three (line 7). While this approach may cause limited overprovisioning, it is acceptable as it remains bounded and cache usage naturally decreases as data is flushed to the servers.

\subsection{Stability Considerations for Decentralized Tuning}
Client-side decentralized control enables scalable adaptivity, but it can also introduce stability risks, including \emph{oscillation} (e.g., reacting to short-lived fluctuations in observed metrics) and \emph{under-utilization} (e.g., many clients simultaneously ``backing off'' under shared contention). CARAT mitigates these risks through several design choices. First, it applies probability-based candidate filtering with a threshold $\tau$ (set to $\tau=0.8$ in our experiments), considering an update only when the model indicates a sufficiently strong likelihood of meaningful improvement, thereby reducing sensitivity to noise and low-confidence changes. Second, CARAT uses conditional score selection with regularization (Algorithm~\ref{alg:calc_score}) to avoid systematically selecting overly conservative configurations when multiple candidates are viable.

CARAT further restricts actuation to bounded, discrete configuration spaces for both RPC and cache parameters, preventing unbounded drift and enabling repeatable behavior. Finally, CARAT follows a two-stage tuning principle: it adjusts RPC parameters periodically during I/O-active phases, while updating cache limits more conservatively at I/O boundaries. This boundary-aligned cache actuation acts as a stabilizing gate, reducing sensitivity to transient bursts and limiting interference with in-flight RPC dynamics. Together, these mechanisms allow CARAT to operate online and re-adjust promptly as conditions change, without inducing prolonged instability or persistent performance degradation.

%% file: sc/4_eval.tex
\subsection{Cluster Setup}
We evaluated \texttt{CARAT} on the publicly available CloudLab platform~\cite{duplyakin2019design} for reproducibility. 
%all evaluation scripts and disk images are published for replication\footnote{Link omitted for anonymity purposes}. 
We did not use proprietary HPC systems because \texttt{CARAT}'s actuation and observability require elevated privileges: updating Lustre client tunables (e.g., RPC window size, RPCs-in-flight, and client cache limits) and accessing certain client-local/low-level counters (e.g., LLITE statistics) typically requires root-level access, which is often unavailable to end users on production systems.
Accordingly, we assume an administrator-managed deployment model in which \texttt{CARAT} runs as a system service on compute nodes (or is executed with the necessary privileges), rather than being invoked directly by individual applications.

Our testbed used ten CloudLab c6525-25g machines (Table~\ref{tab:hardware_spec}). We deployed Lustre 2.15.5 with one dual-purpose node as MGS+MDS, four OSS nodes (each OSS hosting two OSTs mapped to two SSD partitions), and five client nodes. Unless stated otherwise, default striping was used. Each experiment was repeated five times to account for system noise and variability.

\begin{table}[t!]
\footnotesize
\centering
\caption{Hardware Specification for c6525-25g Node}
\begin{tabular}{ll}
\toprule
\textbf{Component} & \textbf{Specification} \\ 
\midrule
CPU   & 16-core AMD 7302P at 3.00GHz \\ 
\midrule
RAM   & 128GB ECC Memory (8x 16 GB 3200MT/s RDIMMs) \\ 
\midrule
Disk  & Two 480 GB 6G SATA SSD \\ 
\midrule
NIC   & Two dual-port 25Gb GB NIC (PCIe v4.0) \\ 
\bottomrule
\end{tabular}
\label{tab:hardware_spec}
\vspace{-1em}
\end{table}

%coupled with version 2.12.5 of the Lustre to build our cluster. We selected three categories of machines from CloudLab: c220g1, c220g2, and c220g5. These machines boast similar hardware specifications, each featuring two Intel Xeon CPUs (16-20 physical cores), DDR4 main memory (128-192 GB), an Intel DC S3500 SATA SSD (480 GB), one to two 10K RPM SAS SFF HDDs (1 TB to 1.2 TB), and a 10Gb dual-port Intel X50-DA2 NIC. For the parallel file system (PFS) shared storage, only SSDs were utilized. The c220g1 and c220g2 machines supported single-client executions, while the c220g5 machines were reserved for multi-client scenarios, accommodating up to five clients. Our server configuration for both single and multi-client executions included one machine dual-functioning as the management server (MGS) and the metadata server (MDS), along with four machines serving as object storage servers (OSS). The MGS and MDS setups each hosted a single target, and each OSS contained two targets, with each target being a storage partition of 200 GB. 

\subsection{Workload Setup \& Data Collection}
\label{subsec:wld}
A core principle of \texttt{CARAT} is the creation of local metrics that effectively capture the influence of I/O patterns on the global storage system. This approach uniquely enables the training of our model using simple training data, while still achieving high accuracy when applied to more complex I/O patterns and realistic HPC executions. To validate this principle, the ML models were trained using the simplest workloads in Filebench~\cite{tarasov2016filebench}: single-stream I/O patterns (processes accessing a single file on a single OST) with varying access patterns (sequential or random) and request sizes (8 KB, 1 MB, 16 MB). We used the following naming convention to refer the used Filebench workload during evaluations: \texttt{{[stream\_type]\allowbreak\_[operation\_type]\allowbreak\_[access\_type]\allowbreak\_[size]}}.
Here, stream\_type indicates either single stream or five streams ($s, f$), operation\_type represents either read or write ($rd, wr$), access\_type denotes either sequential or random ($sq, rn$), and request\_sizes are specified by their actual sizes ($8k, 1m, 16m$).

\textbf{Training Data.} To collect training data, we simulated both \texttt{Read} and \texttt{Write} operations using specified I/O patterns. Each pattern was executed for 300 seconds and repeated 30 times to generate a sufficient number of samples. During these simulations, we probed the system at half-second intervals to gather raw statistics and made random adjustments to the targeted tunable parameters after each probe. These random tuning actions caused the I/O client to experience a variety of RPC communication scenarios, allowing for a thorough evaluation of the designed metrics' effects and providing valuable insights for the machine learning model's training.

\textbf{Evaluation Workloads.} We evaluated the ML models using new, unseen constant workload, dynamically changing workloads, workloads with interferences, and real-world HPC applications. 
%\textit{five-stream} I/O patterns from Filebench (i.e., five independent processes accessing five large files across multiple OSTs)

\begin{comment}
\begin{figure}
    \centering
    \includegraphics[width=0.48\textwidth]{sc/figs/loss_curve.pdf}
    \caption{The learning curve of GBDT read and write models}
    \label{fig:loss_curve}
\end{figure}
\end{comment}

\renewcommand{\arraystretch}{1}
\begin{table}[t!]
\caption{Error Rates of ML Models}
\footnotesize
\centering
\begin{tabular}{ccc}
\toprule
 & Read Model  & Write Model \\
\midrule
SVM~\cite{cortes1995support} & 0.23 & 0.25 \\
\midrule
FC-NN & 0.11 & 0.21 \\
\midrule
RNN~\cite{rumelhart1986learning} & 0.17 & 0.23 \\
\midrule
TCN~\cite{bai2018empirical} & 0.11 & 0.28 \\
\midrule
GBDT~\cite{friedman2001greedy} & \textbf{0.04} & \textbf{0.12} \\
\bottomrule
\end{tabular}
\label{tab:performance}
\vspace{-1em}
\end{table}

\subsection{ML Model Accuracy Evaluation}
\label{sec:ML_eval}
We trained and tuned multiple classifiers on offline data comprising 100{,}730 read-only and 98{,}078 write-only nonzero samples, split 80{:}20 into train/validation. We use \emph{error rate} (i.e., $1{-}$accuracy) as the metric and perform hyperparameter search for each model. Table~\ref{tab:performance} summarizes results. The read/write gap reflects different operation complexity in Lustre (see \S~\ref{subsec:ML_model}).

GBDT~\cite{friedman2001greedy} attains the lowest error for both reads (0.04) and writes (0.12), and we adopt it in \texttt{CARAT}. Its tree-based learning captures nonlinear feature--performance relations well on moderately sized datasets and is robust to noise and sparsity. Neural models (FC-NN, RNN, TCN) can overfit with limited data, particularly TCN on writes, while SVM underfits due to its simpler decision boundary. Beyond aggregate error, our qualitative inspection suggests the read-focused model is generally more accurate and stable than the write-focused model, which is consistent with the greater complexity and variability of write behavior in Lustre. We therefore use the learned model primarily as a lightweight filter to eliminate clearly unpromising configurations; the subsequent probability thresholding and bounded configuration space in the online tuner further reduce the impact of occasional mispredictions and prevent overly aggressive actions.

\subsection{Tuning Evaluations}
\label{sec:tun_eval}
We conducted \textbf{five} sets of evaluations to validate \texttt{CARAT}'s tuning capabilities under diverse I/O patterns and system conditions, summarized as follows:

\begin{itemize}[leftmargin=*]
    \item \textbf{Static workloads:} Firstly, we assessed the \texttt{CARAT} to static I/O patterns. We evaluated both seen and unseen Filebench workloads to verify its effectiveness. (\S\ref{sec:standalone_execution}).
    
    \item \textbf{Dynamic workloads:} Secondly, we demonstrated tuning adaptability of \texttt{CARAT} on dynamically changing I/O patterns. We evaluated how fast and how accurate \texttt{CARAT} reacts to the I/O pattern changes (\S\ref{sec:dynamic_execution}). 
    
    \item \textbf{Independent tuning:} Thirdly, we evaluated how \texttt{CARAT} can make distinct tuning decisions on different I/O clients of the same applications and illustrated whether it will improve the overall I/O performance of the application (\S\ref{sec:independent_execution}).
    
    \item \textbf{External interference:} Fourthly, we evaluated the performance of \texttt{CARAT} when there are interference I/O behaviors in the global storage system, aiming to show the effectiveness of using only local metric to handle global storage system status (\S\ref{sec:intf_execution}).
    
    \item \textbf{Real-world applications:} Lastly, we validated \texttt{CARAT} tuning capabilities on real-world deep learning and HPC applications beyond benchmark workloads. (\S\ref{sec:real_execution}).
\end{itemize}

%Most evaluations, except for cases that involve imbalanced I/Os and external interference, were conducted using a single client. Given \texttt{CARAT}'s decentralized design, the single-client evaluations will effectively demonstrate its ability to adapt to varying I/O patterns. The external interference evaluation (\S\ref{sec:intf_execution}) extended this by running five clients executing distinct workloads under varying interference conditions, showcasing cross-application adaptability.

\begin{figure}[t!]
    \centering
    \includegraphics[width=0.85\columnwidth]{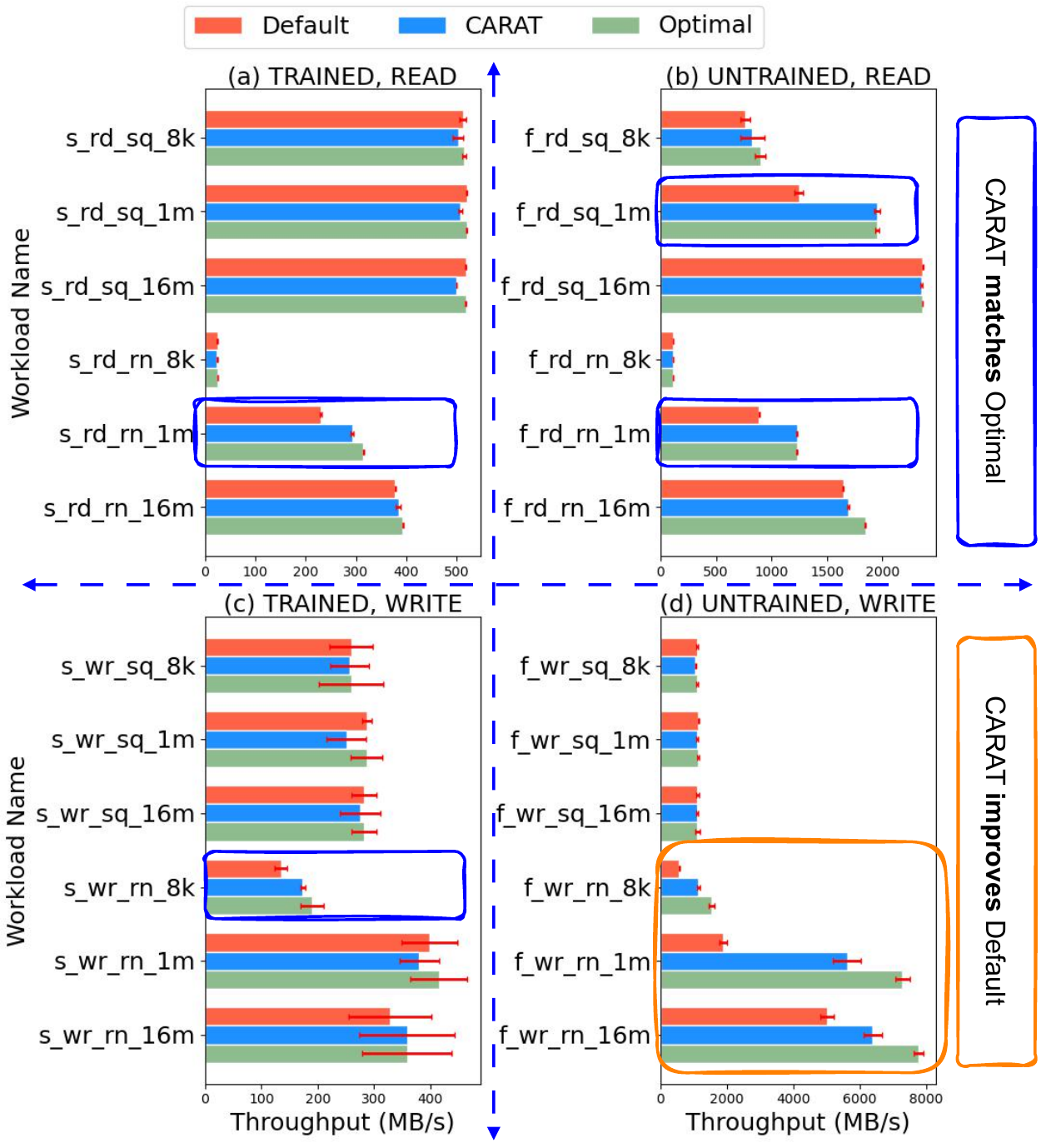}
    \caption{Performance comparisons of tuning static workloads.}
    \label{fig:standalone_test}
    \vspace{-1em}
\end{figure}

\subsection{Tuning Static Workloads}
\label{sec:standalone_execution}

We first evaluated \texttt{CARAT} towards static executions of different Filebench I/O workloads (both seen and unseen I/O patterns)~\cite{tarasov2016filebench}. 
%Specifically, we used the framework to tune parameters while running different Filebench workloads (both seen and unseen)~\cite{tarasov2016filebench}. 
These evaluations focus on testing the framework's performance to static I/O patterns. %thereby validating its capacity to handle a variety of workloads effectively.
%, in addition to the I/O patterns it was trained on
%\textcolor{orange}{[NRT: Nice. To help a reader quickly test against Optimal, could you add an annotation, eg, a black dot for each workload?]}
%on a single client cluster 
For comparison, we ran the workloads under three distinct execution scenarios: \textbf{default}, \textbf{tuning}, and \textbf{optimal}. The default scenario utilized Lustre's default configuration. In the tuning scenario, \texttt{CARAT} dynamically adjusted Lustre's configurations. The optimal scenario involved applying the best possible configuration for each workload, determined through simulations across all possible configurations offline. We executed each workload five times, reporting the average performance alongside the standard deviation as an error bar in our visual results.

Figure~\ref{fig:standalone_test} reports the results. In total, we evaluated 24 different I/O workloads from Filebench. The top row includes all \texttt{Read} workloads and the bottom row includes all \texttt{Write} ones. The left column includes all I/O workloads that were used during training. The right column, however, includes only those workloads that are never seen by the model. Among all the results, we highlight two cases, where \textit{default} and \textit{optimal} configurations have significant performance differences. Among them, the blue rectangles circle cases where \texttt{CARAT} matches the \textit{optimal}; while the yellow ones circle cases where \texttt{CARAT} improves the \textit{default} baseline and results in performance closer to \textit{optimal}.
%\textcolor{orange}{[NRT: Add annotations? (probably already known)]}

From these results, we can observe that across all workloads, \texttt{CARAT} either matched the performance of the \textit{default} configuration (Baseline) within a 10\% variation or significantly enhanced it, achieving up to a \textbf{3-fold} increase in performance by matching the \textit{optimal}. The \textbf{Figure~\ref{fig:standalone_test}(d)} is exceptionally worth noting. For such an 1M write I/O pattern, traditional wisdom is to maximize RPC generation and transfer speed to send more data. However \texttt{CARAT} chooses to reduce RPC generation speed but delivers better performance. We checked the case and noticed that this workload contains heavy in-place updates, hence could leverage write dirty cache if the writes can be buffered in the cache longer. \texttt{CARAT} identifies and effectively exploits this opportunity.

\begin{figure*}[t!]
    \centering
\includegraphics[width=\textwidth]{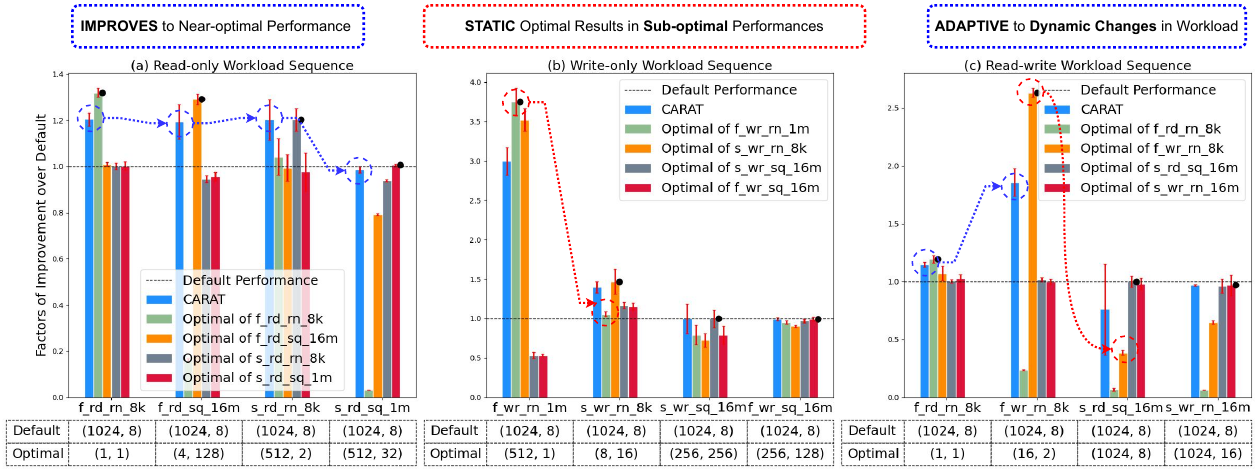}
    \caption{Performance (throughput) comparisons of tuning dynamic workloads. Both \textit{default} and \textit{optimal} configurations are shown under the respective I/O patterns in the following format (`RPC Window Size', `RPCs in Flight'). The black dot represents the \textit{optimal}.}
    \label{fig:dynamic_test}
    \vspace{-1em}
\end{figure*}

\subsection{Tuning Dynamic Workloads}
\label{sec:dynamic_execution}

We evaluated \texttt{CARAT} with dynamically changing I/O patterns. In this scenario, \texttt{CARAT} is effective because it captures the changes instantly and tunes each I/O client independently and rapidly. 
%The objective of conducting dynamic executions of workloads was to demonstrate \texttt{CARAT}'s ability to adapt in real time to varying I/O patterns and to validate its proficiency in dynamic parameter tuning. 
In these evaluations, we created dynamically changing I/O patterns by running a sequence of Filebench workloads. To be general, we explored three sequences of workloads: a sequence of various read workloads, a sequence of various write workloads, and a sequence of mixed read and write workloads. Each sequence includes four workloads. After 300 seconds of running a particular workload, the execution would transition to the next workload. 

Similar to the previous scenario, we compared both \textbf{default} and \textbf{optimal}. However, the optimal is no longer one configuration. Instead, we evaluate on optimal of each workload of the sequence. We named them based on the workload name and plots their performance individually as shown in Figure~\ref{fig:dynamic_test}. 
Here, subplots (a, b, c) show the performance of three sequences (Read-only, Write-only, and Mixed). The four groups along the $x$-axis are the four Filebench workloads that were executed in order in the sequence. 
The bars show the performance of \texttt{CARAT} and different \textit{optimal} configurations. They are all normalized based on the \textit{default} configuration (1.0 in $y$-axis). Please note that \texttt{CARAT} is always comparing with the optimal configuration of each I/O workload.
%conducted these sequences under both the default configurations and the optimal configurations tailored for each workload within a sequence.

The results clearly illustrate the \texttt{CARAT}'s capability and advantages to adjust continuously to the changing I/O patterns. Across all the cases, \texttt{CARAT} is able to adapt to the new I/O pattern and achieve near-optimal performance. We observed overall improvements up to \textbf{3$\times$} better than the baseline for the I/O patterns in the sequence. %than the default performance with our tuning framework. 
The individual \textit{optimal} static configurations experienced much worse performance as the I/O pattern changes.

\subsection{Performance of Independent Tuning}
\label{sec:independent_execution}
We assess \texttt{CARAT} when an application’s processes exhibit different I/O behaviors. Table~\ref{tab:independent} shows two processes on a single client issuing simultaneous 8\,KB sequential write (Process-1) and read (Process-2) in a file-per-process mode via different I/O clients. We compare \texttt{CARAT}’s online adaptation against three static choices: the Lustre default and two fixed settings (\texttt{Optimal-1}, \texttt{Optimal-2}) that correspond to the most frequently selected per-process configurations observed during tuning; tuples denote (\textit{RPC window pages}, \textit{RPCs in flight}). \texttt{CARAT} consistently delivers higher throughput for both processes, demonstrating that per-client dynamic tuning outperforms static “optimal” settings as conditions evolve across co-running read/write streams.

\subsection{Tuning under External Interference}
\label{sec:intf_execution}
We evaluate \texttt{CARAT} under server-side contention created by concurrent clients, testing whether client-local decisions generalize to shared, interfering environments. When multiple clients target overlapping OSTs, queueing rises and service rate fluctuates. Bursty writers inject many under-filled RPCs that inflate per-RPC fixed costs; competing readers see elevated per-page latency and sporadic channel idling; mixed read–write phases exacerbate head-of-line blocking and cache-pressure–induced flush spurts. We induce these effects by running five distinct workloads from five clients against overlapping OSTs in three scenarios: all-read, all-write, and mixed set of five workloads.

\begin{table}[t!]
\caption{Independent Tuning Executions}
\footnotesize
\centering
\begin{tabular}{ccccc}
\toprule
& Default & Optimal-1 & Optimal-2 & \textbf{\texttt{CARAT}}\\
& (MB/s) & (MB/s) & (MB/s) & \textbf{(MB/s)}\\
\midrule
Process-1 & 76.5 & 73.8 & 77.8 & \textbf{93.6}\\
\midrule
Process-2 & 14.4 & 25.1 & 26.1 & \textbf{58.1}\\
\midrule
Configuration & (1024, 8) & (1024, 256) & (64, 256) & \textbf{Dynamic}\\
\bottomrule
\end{tabular}
\label{tab:independent}
\vspace{-1em}
\end{table}

\begin{table}[t!]
\centering
\caption{External Interference Executions}
\footnotesize
\begin{tabular}{ccc}
\toprule
\textbf{External Interference Type} & \multicolumn{2}{c}{\textbf{Aggregated Cluster Throughput}} \\
\cmidrule{2-3}
 & Default (MB/s) & \textbf{\texttt{CARAT} (MB/s)} \\
\midrule
All Read Workloads & 1223.04 & \textbf{1403.80} \\
\midrule
All Write Workloads & 2445.64 & \textbf{3604.38} \\
\midrule
Mix of Read--Write Workloads & 1030.96 & \textbf{3057.08} \\
\bottomrule
\end{tabular}
\label{tab:external_intf}
\vspace{-1em}
\end{table}

\texttt{CARAT} reacts to these conditions using only local signals: it trims in-flight concurrency when per-page RPC latency climbs and channel utilization stalls, enlarges the RPC window to reduce small-RPC bursts once contention appears, and on write-heavy paths, raises dirty-cache headroom for active clients to preserve coalescing while keeping inactive ones at minimal cache. As contention eases, it restores concurrency to re-expose parallelism. This combination suppresses premature dispatch, stabilizes channel utilization, and smooths server queues without global coordination, improving aggregate efficiency across clients. As shown in Table~\ref{tab:external_intf}, aggregate cluster throughput improves over the default by \textbf{15\%} (all-read), \textbf{1.47$\times$} (all-write), and up to \textbf{3.0$\times$} (mixed), reflecting better extent filling, right-sized concurrency, and adaptive cache allocation under interference.

\subsection{Tuning Real-world Applications}
\label{sec:real_execution}
% This evaluation aimed to assess \texttt{CARAT}'s ability to adapt to the actual I/O patterns of real applications despite being trained on only a limited set of benchmark data. We focused on both emerging applications in HPC systems and those commonly found in traditional HPC environment.
We evaluate \texttt{CARAT} on real applications to test generalization beyond its benchmark-trained models, covering both emerging deep learning (DL) workloads and traditional HPC applications. Importantly, these application traces and kernels are not used in training; thus, this experiment evaluates whether the client-local metrics captured by \texttt{CARAT} can guide tuning decisions even when the exact application access pattern was not seen during offline model construction.

\begin{figure}[t!]
    \centering
    \includegraphics[width=0.80\columnwidth]{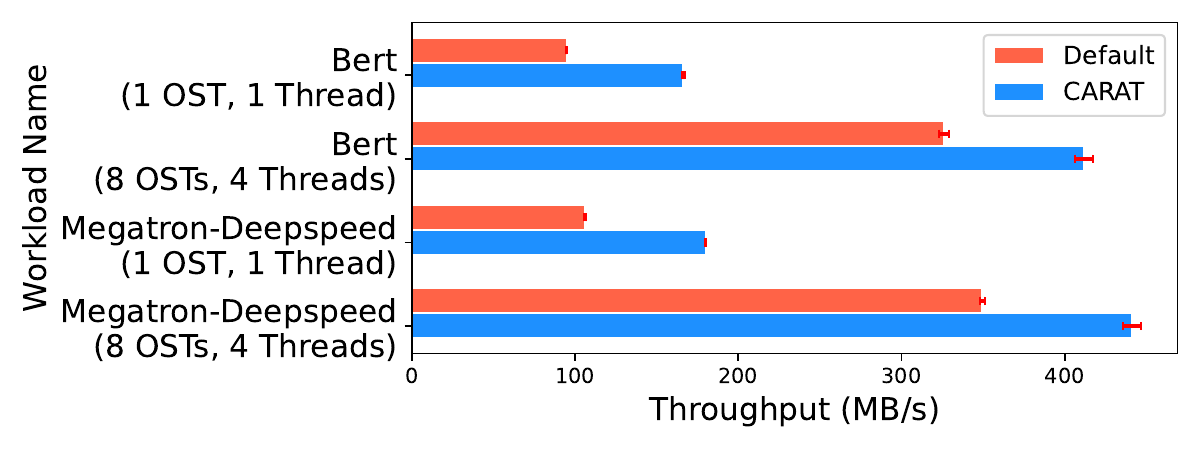}
    % \caption{Performance comparisons of DL applications.}
    \caption{DL I/O kernels (BERT~\cite{devlin2018bert}, Megatron/DeepSpeed~\cite{shoeybi2019megatron,rasley2020deepspeed}) via DLIO~\cite{devarajan2021dlio}: throughput vs. default.}
    \label{fig:dlio_test}
    \vspace{-1em}
\end{figure}

% We first evaluated the growing presence of deep learning (DL) applications within the HPC ecosystem. Their unique I/O patterns have placed lots of challenges to traditional I/O optimization methods~\cite{lewis2025machine}. We examined two distinct deep learning I/O kernels: one representing the BERT natural language processing model~\cite{devlin2018bert} and the other a Deepspeed~\cite{rasley2020deepspeed} version of NVIDIA’s Megatron~\cite{shoeybi2019megatron}. These I/O kernels were evaluated using the deep learning I/O benchmark (DLIO)~\cite{devarajan2021dlio} across different utilization of storage targets (OSTs) and varying numbers of threads. The evaluations in Figure~\ref{fig:dlio_test} revealed a notable improvement over the default configurations, with enhancements reaching as high as \textbf{1.75 times} the baseline performance. This finding highlights the framework's capacity for effectively tuning real-world applications.
DL workloads pose distinctive I/O behaviors that challenge traditional tuning~\cite{lewis2025machine}. Modern DL training is dominated by small, sample-oriented reads spread over numerous files, plus per-epoch shuffling and multi-threaded prefetch that create short, bursty I/O phases. These traits fragment RPC extents, trigger under-filled requests, and intermittently overwhelm or idle RPC channels. Using DLIO~\cite{devarajan2021dlio}, we evaluate BERT~\cite{devlin2018bert} and Megatron with DeepSpeed~\cite{shoeybi2019megatron,rasley2020deepspeed} across varying OST utilization and thread counts. 
%As shown in Fig.~\ref{fig:dlio_test}, \texttt{CARAT} improves throughput over default by up to \textbf{1.75$\times$}, demonstrating effective adaptation to real DL I/O. 
\texttt{CARAT} adapts the RPC window to sustain channel utilization under small/unaligned reads, tempers in-flight concurrency under momentary contention, and raises dirty-cache headroom to coalesce bursts. This behavior is consistent with the design goal of filtering out low-confidence changes and operating within a bounded configuration space: rather than continuously exploring, \texttt{CARAT} applies only high-confidence adjustments that improve the likelihood of maintaining steady effective concurrency during bursty phases. In our DLIO runs, this combination reduced premature small-RPCs during bursts and preserved coalescing when sequential access emerged, enabling the observed \textbf{up to 1.75$\times$} speedup over the default as shown in Fig.~\ref{fig:dlio_test}. Since the total data volume is fixed per run, these throughput gains translate into corresponding reductions in the I/O phase time for the DL kernels.

\begin{table}[t!]
\centering
\caption{HPC Scientific Application Executions}
\footnotesize
\begin{tabular}{ccc}
\toprule
\textbf{Scientific HPC Applications} & Default(MB/s) & \textbf{\texttt{CARAT}(MB/s)}\\
\midrule
VPIC-IO (3D array write)	& 459.7	& \textbf{514.6}\\
\midrule
BDCATS-IO (3D array read)	& 398.2	& \textbf{405.3}\\
\bottomrule
\end{tabular}
\label{tab:hpc_app}
\vspace{-1em}
\end{table}

% We evaluated the framework's effectiveness using traditional HPC applications. We employed H5bench~\cite{h5bench} benchmark, which features I/O kernels modeled after real-world applications: the \texttt{VPIC-IO} kernel for writing data, based on a particle physics simulation~\cite{byna2012parallel}, and the \texttt{BDCATS-IO} kernel for reading that data, based on a big data clustering algorithm. As shown in Table~\ref{tab:hpc_app}, the framework delivers either on-par or improved performance compared to the default, demonstrating its applicability to real HPC applications.
For traditional HPC, we use H5bench~\cite{h5bench}: \texttt{VPIC-IO} (write) derived from a particle-physics code~\cite{byna2012parallel} and \texttt{BDCATS-IO} (read) based on a big-data clustering workflow. Table~\ref{tab:hpc_app} shows \texttt{CARAT} is on-par or better than default, confirming applicability to classic scientific workloads. We note that improvements on these kernels can be smaller than on DLIO because their I/O behavior is more sequential and regular, for which the Lustre client default RPC and caching settings are often already near-optimal. As a result, there is limited headroom for additional gains from client-side tuning; in these cases, \texttt{CARAT} typically either makes minor adjustments when it identifies a clear opportunity or remains close to the default settings.

%The results in Table~\ref{tab:hpc_app} showed performance on par with the optimal configuration, corroborating our previous findings that, \texttt{CARAT} can effectively deliver near-optimal performance. {\color{red}Table 5 should be dropped and becomes a sentence in this paragraph. We observe that \texttt{CARAT} always achieve optimal throughput easily.}
%\textcolor{orange}{[NRT: Table 5 is overhead... I think the H5bench results are missing?]}

\begin{comment}
\setlength{\belowdisplayskip}{-10pt} % adjust the space after displays
\begin{table}[htbp]
\centering
\caption{HPC Scientific Application Executions}
\footnotesize
\begin{tabular}{ccc}

\toprule
\textbf{Scientific} & \textbf{Optimal} & \textbf{\texttt{CARAT}'s}\\
\textbf{HPC} & \textbf{Throughput} & \textbf{Throughput}\\
\textbf{Applications} & \textbf{(MB/s)} & \textbf{(MB/s)}\\
\midrule
VPIC-IO (1D array write)	& 327.2	& 321.6\\
\midrule
VPIC-IO (2D array write)	& 319.4	& 317.7\\
\midrule
VPIC-IO (3D array write)	& 331.1	& 326.1\\
\midrule
BDCATS-IO (partial read)	& 441.3	& 436.1\\
\midrule
BDCATS-IO (strided read)	& 455.5	& 455.5\\
\midrule
BDCATS-IO (full read)	    & 463.3	& 463.1\\
\bottomrule

\end{tabular}
\label{tab:hpc_app}
\end{table}
\end{comment}

\subsection{\texttt{CARAT} Overhead}

% \texttt{CARAT} operates independently on each client, to achieve efficiency in both storage and resource usage. It requires minimal storage space, maintaining two snapshots in the I/O client's main memory at any time. Additionally, because \texttt{CARAT}'s I/O operations are limited to collecting statistics from the local system without directly interacting with the remote shared storage systems, it avoids adding extra I/O and network overhead to the PFS.
\texttt{CARAT} runs independently per client, maintains only two in-memory snapshots, and collects client-local statistics without issuing remote I/O, thereby avoiding extra PFS or network traffic. At runtime, each client executes the same control loop locally: it probes metrics, selects the appropriate (read- or write-focused) model based on the observed I/O mix for that interval, performs inference once (i.e., it does not run both models), and applies an update only when the candidate passes the probability threshold; otherwise it retains the current setting.

% Table~\ref{tab:overhead} reports the average time taken for key operations: generating snapshots from probing, performing inference using the trained model, and completing the tuning process, which includes inference, identifying the optimal configuration, and applying it for both read- and write-centric tuning scenarios across each I/O client. These were measured on the CloudLab machines. Notably, probing, inference, and tuning for each I/O Client can be done in parallel, meaning the overhead does not scale with the number of ranks or I/O clients or computing nodes. This demonstrates \texttt{CARAT}'s strong scalability and minimal resource impact.
Table~\ref{tab:overhead} reports per-client averages (CloudLab) for snapshot creation, model inference, and end-to-end tuning (inference + selection + apply) under read- and write-centric workloads. Probing, inference, and tuning execute in parallel across clients, so overhead does not grow with ranks/clients/nodes, demonstrating strong scalability. This addresses the concern that frequent invocation could amplify overhead at scale: the overhead is incurred independently per client and does not introduce centralized bottlenecks. Moreover, the end-to-end tuning cost reported in Table~\ref{tab:overhead} remains well below the probing period used in our experiments, leaving slack for stable operation even when tuning is evaluated frequently.

% Note that, for these evaluations, we used a very short probing interval of 0.5 seconds to probe the local file system, which was made possible by the use of local-only metrics and autonomous execution. The probing interval can be adjusted, allowing users to customize the framework's operation to suit their specific needs.
We used a short 0.5\,s probing interval made possible due to usage of local-only metrics; this interval is user-configurable. The probing interval presents a responsiveness--stability tradeoff: shorter intervals can react more quickly to phase changes, while longer intervals reduce control-loop activity and further damp sensitivity to transient variations. In practice, \texttt{CARAT} can be run at coarser intervals for less I/O-intensive workloads or when a more conservative tuning cadence is desired.

\begin{table}[t!]
\caption{\texttt{CARAT} Overheads per I/O Client}
\footnotesize
\centering
\begin{tabular}{cccc}
\toprule
 & \textbf{Snapshot} &  & \textbf{End-to-end}\\
\textbf{Operation} & \textbf{Creation} & \textbf{Inference} & \textbf{Tuning}\\
\textbf{Type} & \textbf{Time (ms)} & \textbf{Time (ms)} & \textbf{Time (ms)}\\
\midrule
Read & 0.33 & 10.06 & 24.64\\
\midrule
Write & 0.85 & 13.51 & 28.82\\
\bottomrule
\end{tabular}
\label{tab:overhead}
\vspace{-1em}
\end{table}

%% file: sc/5_related.tex
Auto-tuning for HPC I/O spans a large, coupled search space and is algorithmically hard~\cite{dorier2022hpc,bernstein2002complexity}. Prior systems therefore adopt approximations along two main lines. 

\textbf{Heuristic/Rule-based tuning.} Rule-driven methods encode expert knowledge or lightweight policies for specific knobs and patterns. Li et al.~\cite{li2015ascar} propose a rule-based strategy; TAPP-IO~\cite{neuwirth2017automatic} and DCA-IO~\cite{kim2019dca} tune Lustre striping (size/count) using heuristic and log-informed rules, and IOPathTune~\cite{rashid2023iopathtune} exploits client-side metrics for adaptive tuning. While efficient, such approaches struggle to generalize across diverse and time-varying I/O behaviors, and often lack a mechanism-aware way to reconcile RPC coalescing, concurrency, and cache pressure that arise prominently in PFS clients.

\textbf{Black-box optimization and learning.} A complementary line that models performance and searches configurations with black-box optimizers. Cao et al.~\cite{cao2018towards} compare auto-tuning strategies across file systems; SAPPHIRE~\cite{lyu2020sapphire} applies Bayesian optimization; Behzad et al.~\cite{behzad2019optimizing} couple nonlinear regression with genetic algorithms. Recent systems like CAPES~\cite{li2017capes}, AIOC2~\cite{cheng2021aioc2}, and Magpie~\cite{zhu2022magpie} use DRL to learn end-to-end policies. These methods can be effective but typically depend on global instrumentation and aggregation, incurring nontrivial data-collection, training, and coordination overheads that impede online, per-client adaptation at scale.

\textbf{Positioning of \texttt{CARAT}.} In contrast, \texttt{CARAT} narrows the scope to \emph{client-local} observability and actuation. It uses low-level metrics that directly reflect RPC extent utilization, in-flight concurrency, and dirty-cache dynamics, and applies lightweight ML to make online decisions per client without central coordination. This design sidesteps global telemetry cost while addressing the core client-side trade-offs that prior heuristics and global black-box methods only indirectly capture, enabling scalable RPC+cache co-tuning under dynamic workloads.

%% file: sc/6_conclude.tex
We presented \texttt{CARAT}, a ML-guided system for \emph{client-side RPC and cache co-tuning} that operates online using only client-local signals. \texttt{CARAT} decouples fast-acting RPC control from slower cache-limit adjustments via a two-stage tuning, and selects configurations with a conditional score–greedy policy informed by lightweight ML models. Implemented atop Lustre, \texttt{CARAT} improves throughput by up to \textbf{3$\times$} across benchmarks and real applications (DLIO kernels, H5bench), while incurring negligible per-client overhead and requiring no application changes or cluster-wide coordination.
Our results indicate that client-side observability and actuation can effectively manage extent underutilization, congestion, and cache pressure in production-style settings. In future, we plan to broaden the tunable surface (e.g., additional client/runtime knobs)and extend the approach to metadata paths and other RPC-based PFSes. We believe this direction enables more scalable, timely, and tunable storage for data-intensive HPC workloads.